\documentclass[graybox]{SNmult}

\usepackage{url}
\usepackage{hyperref}
\usepackage[square,numbers]{natbib}

\usepackage{type1cm}        
\usepackage{makeidx}         
\usepackage{graphicx}        
\usepackage{multicol}        
\usepackage[bottom]{footmisc}

\usepackage{newtxtext}       %
\usepackage[varvw]{newtxmath}       

\makeindex             

\begin{document}

\title*{Explainable AI through the Lens of Material Agency: Enabling Musical Interface Design with Neural Audio Models}
\titlerunning{Explainable AI through the Lens of Material Agency} 


\author{Shuoyang Jasper Zheng, Anna Xambó Sedó, and Nick Bryan-Kinns}
\authorrunning{SJ. Zheng et al.} 

\institute{Shuoyang Jasper Zheng \at Centre for Digital Music, Queen Mary University of London, London, United Kingdom \\ \email{shuoyang.zheng@qmul.ac.uk}
\and Anna Xambó Sedó \at Centre for Digital Music, Queen Mary University of London, London, United Kingdom
\and Nick Bryan-Kinns \at Creative Computing Institute, University of the Arts London, London, United Kingdom
}
%
%
\maketitle

Under review for ``Explainable AI for the Arts'' (N. Bryan-Kinns, Ed.), Springer.
\vspace{5pt}

\abstract{Recent work in Human-Computer Interaction (HCI) increasingly treats AI models as design materials that have distinctive computational properties to shape design artifacts. Artists learn to work with the model ``at play'' to explore their emerging properties. The aim of explainability, in this view, is to make visible a crafting and hacking space to enable sustained creative practices with AI. In this chapter, we propose \textit{material explainability} as a range of activities and artifacts that transform AI models into accessible and inclusive design materials in the workspace of artists, designers, and makers. We present a case study of building a repository of resources to enable artistic explorations of neural audio models in New Interfaces for Musical Expression (NIME) design. Reflecting on our community-building journey and the making of a collection of musical interface designs with a group of artists, we raise three recommendations on enabling the exploration of AI as materials in artistic practices to inspire future XAI design for artists.
}

\section{Introduction}
\label{sec_intro}

The materiality of interaction design, in Human-Computer Interaction (HCI), emphasizes that the emergent characteristics of design materials can drive the design process \citep{tholander_understanding_2012}. Competent designers learn a rich \textit{material understanding} of the grain, resistances, and propensities of materials in their workspace through hands-on practices, and develop technical know-how to work with them \citep{wiberg_material-centered_2018}. In recent discourse in AI and HCI, a trend of acknowledging AI models as design materials is taking place \citep{scurto_prototyping_2021}. Technological advances are thereby situated in creative practices in which designers probe, configure, and appropriate them as materials to unpack their creative possibilities \citep{caramiaux_explorers_2022}, and craft interactive experiences rather than solely focusing on engineering functionalities. 

This material-oriented perspective offers a practice-based lens to XAIxArts, through which explainability can be conceived as working with AIs as materials to understand them ``at play'', through probing, configuring, and tweaking \citep{benjamin_machine_2021}. In this way, the audience for explainability is shifted to artists, designers, and makers, and the aim is to make visible the crafting and hacking space for practice-based exploration with AIs. Therefore, we propose \textbf{material explainability} -- that is, a range of activities or artifacts that aim to transform AI models into accessible and inclusive design materials, to enable sustained creative practices with AI materials. 

In this chapter, we discuss the intersection of explainability and the material-oriented engagement with AI models in creative practices. We reflect on a case study of New Interfaces for Musical Expression (NIME) designs with AI models as materials. The case study focuses on a specific domain of machine learning that relies on explainability -- the latent space, data representations learned by AI models that have typically been treated as a ``black box'' that lacks explainability. We show how we build a repository of resources, including a software toolkit, documentation, tutorials, and a shared space for projects to bring latent space into the workspace of artists, designers, and makers, and how they explore the materiality aspect of latent space to craft musical interfaces.

\section{Towards Explainability for Accessible and Inclusive AI Materials}
\label{sec_related}

We take the broader view of XAIxArts to refer to explainability as understanding how to work with AI models \textit{in action} in artistic contexts \citep{wilson_embodied_2024, zheng_mapping_2024}. In literature such as XAI in large language models \citep{ehsan_explainable_2024}, this way of gaining functional and operational understanding of the model, typically referred to as \textit{around the edges} explanations, aims to ``provide insights into a black box that can afford a wider range of actions'' \citep[p.~8]{ehsan_explainable_2024}. In artistic practices with AIs, a similar ``understanding models in action'' approach has been widely adopted. For instance, dancers exploring the model's parameter space through embodied movements to relate its outputs to their actions \citep{nabi_embodied_2024}, musicians creating annotations for sound-producing gestures to play the model as a musical instrument \citep{zheng_exploring_2025}, or visual artists engaging in technical practice with the model to discover ways of sculpting artistic expressions \citep{cole_kisscrash_2023}. A commonality, as described by Scurto et al. \citep{scurto_prototyping_2021} using the term ``crafting machine learning'', is that artists seek in-depth engagement with their AI models to discover practical necessities of how to work with them, beyond a technical XAI lens. With this view of craft and design, and the iterative cycles of probing, tuning, and programming models as computational material, it is reasonable to look at the ``material turn'' in HCI to unpack the role of explainability.

\subsection{The Material Turn and Material Agency}
\label{sec_material_hci}

There has been a ``material turn'' in HCI and interaction design research over the past fifteen years \citep{giaccardi_foundations_2015} -- that is, a material understanding of how distinctive properties of information technologies shape interaction experience and possibility. Influential characterizations of this material lens have guided research such as \textit{computational composites} \citep{vallgarda_material_2016} which treat code, electronics, and physical matter as tightly coupled composites, and \textit{material-centered interaction design} \citep{wiberg_material-centered_2018} in which designers learn a repertoire of materials and their grain, resistances, and propensities through hands-on work, in a similar way as they would work with wood, textiles, or sound. HCI researchers use \textit{material agency} to describe how emerging properties of materials actively contribute to the way a design activity unfolds. In this view, agency is ``something that emerges and is dynamically enacted through practice'' \citep[p.~2506]{tholander_understanding_2012}, rather than as a pre-existing set of affordances of a particular actor \citep{bown_understanding_2009}. The notion of \emph{intra-action} \citep{barad_meeting_2007} is widely adopted to question the extreme localization of agency -- either material dominates the human or human dominates the material -- instead, musical interactions are co-produced in the designer's entangled practice with their materials \citep{morrison_entangling_2024}.

In recent discourse of AI in interaction design, a similar trend of acknowledging the computational properties of AI as materials is taking place, which we describe below.

\subsection{AI as Design Material}
\label{sec_material_xai}

Scurto et al. \citep{scurto_prototyping_2021} introduce elements of Machine Learning (ML) that constitute its materiality, including \textit{techniques, data, models, and algorithms}. They argue that art-based ML crafting can take these technical elements as an entry point for design, and emphasize on ``the experience of raw computational properties through incremental crafting'' \citep[p.~2022]{scurto_prototyping_2021} rather than solely on engineering functionalities. The materiality of AI is not a singular form of infrastructure or software. Instead, it extends to a network of actors beyond just artists, audience, and the models \citep{xambo_humanmachine_2024}. 

Therefore, the material agency of AI systems in art and design has been described as a \textit{heterogeneous assemblage} that involves a range of ``technically, materially, socially, and temporally'' distributed infrastructures \citep{gioti_composing_2022}. To be clear, one strand of research in HCI on human-AI co-creation focuses on explicitly assigning agency to either the AI or human \citep{zhang_exploring_2025}. This contrasts the intra-action relationship in which AI models are understood as material rather than co-creator \citep{tholander_understanding_2012}, and the material agency only emerges when being surfaced, negotiated, and repurposed. Acknowledging the material agency of AI models means attending to how their affordances, tendencies, and constraints participate in shaping the artifacts and aesthetic commitments, and being able to design, craft, and refine them across time \citep{scurto_prototyping_2021}.

Repositioning AI in this material agency perspective implies that its characteristic behaviors are not merely technical constraints to be overcome, but affordances that designers must learn to work with through practice \citep{yang_investigating_2018}. In this respect, Pelinski et al. \citep[p.~10]{pelinski_ways_2025} propose \textit{technical practice research} to account for the material agency in works that are surrounded by technical development:

\begin{quotation}
We often do exist in a more dialogic process with our technical practice, in which we converse with our materials, and as such, finding space for practice research might allow us to articulate knowledge about how (rather than what) we do.
\end{quotation}

Technical practice research argues that the material-oriented perspective can be neglected when knowledge production is driven by searching for insights and achieving goals. In practice, the real-time and reflexive accounts during the technical development phase play an important role in shedding light on the temporal emergence of material agency \citep{pelinski_ways_2025}. The issue of linear problem-solving narratives in practice research mirrors the challenge of XAI in the Arts \citep{bryan-kinns_xaixarts_2025}. In particular, technical explanations are typically framed as goal and productivity oriented, raising the barrier to appropriating them for creative and artistic practices \citep{bryan-kinns_reducing_2024}.

\subsection{Accessible and Inclusive XAI Materials}
\label{sec_inclusive}

In the landscape of XAIxArts, attention has been drawn to the accessible, inclusive, interdisciplinary practice with AI \citep{bryan-kinns_xaixarts_2025}. This echoes the interest in diversifying dissemination methods in HCI \citep{sturdee_diversifying_2024}, which argues that traditional academic publications as a way of sharing research findings can exclude wider audiences. Yoo et al. \citep{yoo_translating_2025} propose \textit{Alternative Research Outcomes} to challenge the one-size-fits-all research dissemination strategies. They explore situated and dedicated research activities or artifacts, such as documentaries, public exhibitions, multimedia artifacts, as ``new endpoints to translate, communicate, or disseminate research insights in an accessible and engaging form tailored to the intended audience'' \citep{yoo_translating_2025}. 

Practitioners in the XAIxArts community also adopt the view of making explainability engaging to audiences with a range of technical literacy. This is reflected as an \textit{AI as Material} perspective \citep{bryan-kinns_reflections_2024} in which the AI models themselves can be artworks. Artist-led practices such as probing model behaviors, exposing bias, and re-creating glitches emerge to enable engagement with a wider audience. In this way, research insights on model explainability are made accessible through alternative research outcomes such as exhibitions, interactive experiences, or artistic installations.

The XAIxArts discourse on exploring the AI itself as artistic pieces poses open questions on where the creative agency resides \citep[p.~45]{bryan-kinns_reflections_2024}, aiming to answer ``who is responsible for the explanation: the AI model, the artist, or the audience?'' Here, we suggest that it is desirable to consider the dissemination of models into the hands of artists. Therefore, the \textbf{material explainability} aims to answer how to enable artists to develop technical know-how and technical practices with AI materials. A similar notion of material explainability was raised in the XAIxArts Manifesto \citep{bryan-kinns_xaixarts_2025}, which calls for ``demystifying the computational materiality of AI in accessible languages and engaging formats'' for inclusive dissemination of AI as material. 

Following this call for material explainability in a concrete creative-technical setting, we focus on neural audio models in musical interface design, where AI models are both a key site of creative potential and a barrier to hands-on engagement.

\subsection{Neural Audio Models in New Interfaces for Musical Expression}
\label{sec_neural}

To unpack this theme and illustrate the role of explainability in exploring the materiality of AI, we look at a specific domain of generative AI for sound synthesis that is typically considered to lack explainability -- the latent space. In AI for sound analysis and synthesis, the latent space is a compact multi-dimensional map learned by neural audio models. A typical implementation of neural audio models is an autoencoder, in which an encoder compresses audio signals into a latent space, and a decoder reconstructs audio from latent representations. This method of learning audio data allows them to capture complex timbral and spectral characteristics that are difficult to specify explicitly. We direct intrigued readers to the chapter by Yee-King \citep{yee-king_latent_2022} for a comprehensive and inclusive explanation of using latent spaces in creative practices.

In New Interfaces for Musical Expression (NIME), neural audio models open new opportunities by repositioning sound synthesis as a matter of interacting with the learned latent representations. For instance, the \textit{latent space walk} approach of mapping user inputs into latent spaces enables designers to craft novel musical instruments \citep{privato_stacco_2024, zheng_exploring_2025}. This shifts neural audio models from a technical artifact into design materials that can be explored, used, and appropriated. 

Despite the creative possibilities it opens, one of the long-standing criticisms of latent spaces is the absence of easy-to-understand musical attributes encoded across all dimensions. Consequently, the ways of navigating the latent space and relating the sound outputs rely on the designer's sense-making. Therefore, to explore neural audio models in NIME design, designers seek to understand the model’s computational properties through iterative exploration and configuration. However, technical AI research is typically positioned outside creative coding and audio production environments, and limited learning resources raise the barrier to hands-on engagement with the model \citep{clarke_longevity_2025}. In the next section, we present our journey of transforming neural audio models into situated and easy-to-integrate materials in the hands of artists and designers.

\section{Crafting Musical Interactions with AI Audio Latent Space}
\label{sec_crafting}

This section describes a case study of building a community of resources, including software tools, documentation, tutorials, and demo projects, to turn neural audio models into explainable materials for designers. We introduce the \textit{Latent Terrain} toolkit, a package of Max/MSP objects to build customized interactive sound spaces for neural audio autoencoders. We acknowledge that the appropriation of technical artifacts can be open-ended and cannot be contained by a singular effort. Therefore, we hope to use our project only as a heuristic for further work, rather than an ultimate solution.

\subsection{Description of the Research Project}

\begin{figure}[t]
\sidecaption
\includegraphics{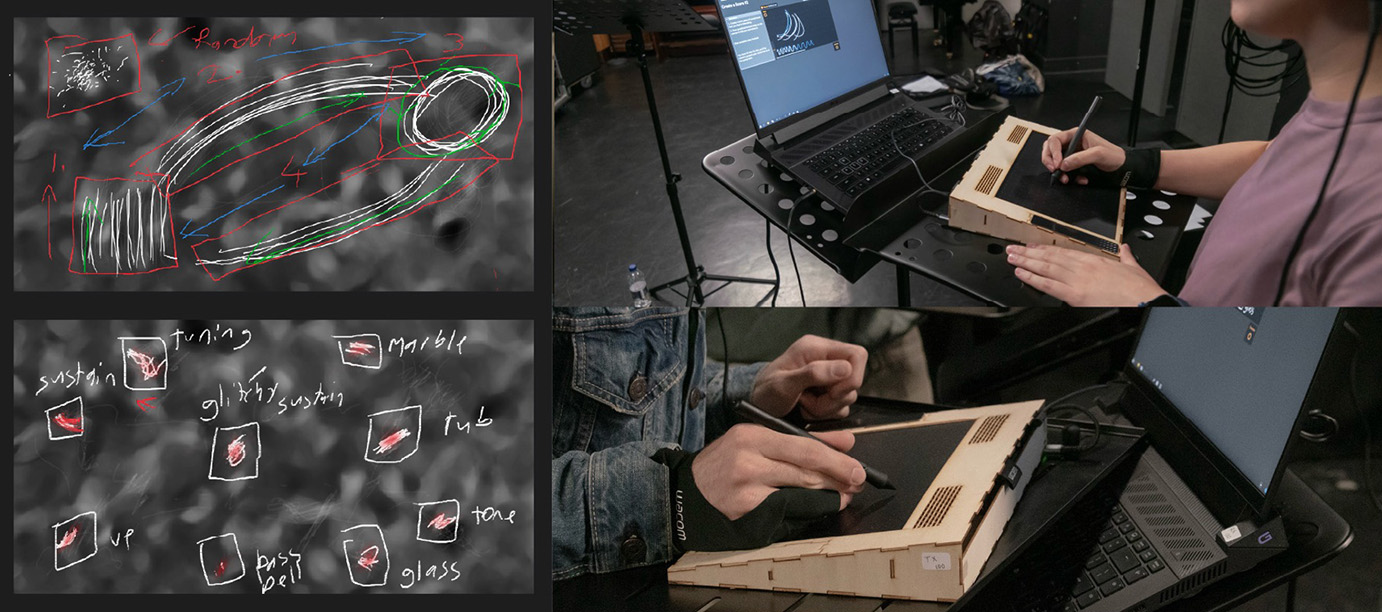}
\caption{We held a series of workshops in which musicians were invited to explore the DMI probe (Right). They were tasked to create musical scores to re-enact sound pieces (Left).}
\label{fig_workshop}
\end{figure}

\textit{Latent Terrain} is the output of the first author's doctoral research project conducted between 2024 and 2025 \citep{zheng_exploring_2025}, at the intersection of NIME, XAI, and HCI. The research looked at the embodied music cognition aspect of movement-sound interaction in audio latent spaces, investigating how musicians explore the latent space as a musical instrument and the performing techniques that emerged. A full technical account of the project is beyond the scope of this paper, we therefore invite readers to our previous work for a detailed journey of the research  \citep{zheng_exploring_2025}. A brief narrative of the journey includes:

\begin{enumerate}
  \item The research took the method of Digital Musical Instruments (DMIs) as probes \citep{jack_digital_2020}, in which musical instruments are built to serve as a means to explore musicians' subjective experience.
  \item We crafted a DMI with a stylus and tablet interface (shown in Figure~\ref{fig_workshop}) that allows musicians to navigate the latent space by operating the stylus, and hear the real-time sound synthesized by the autoencoder.
  \item \textit{Latent Terrain} was developed as a deep learning method to interactively map the latent space of an autoencoder to a 2-dimensional control space, using a customized audio corpora. To facilitate rapid prototyping and testing of the mapping, we developed a toolkit in the Max/MSP visual programming environment that allows us to integrate the Latent Terrain model and the autoencoder together into the DMI.
\end{enumerate}

\subsection{Bridging Computational Materiality to Accessible Platforms}

Initially, the Latent Terrain toolkit was published as supplementary material in a research article \citep{zheng_exploring_2025}. As the project progressed, the intended audience shifted from academic researchers to broader DMI makers in the field of AI audio synthesis. This motivated us to maintain the toolkit as a public Max/MSP (a visual programming language for music and multimedia) package \footnote{Our Latent Terrain Toolkit: \url{https://jasper-zheng.github.io/nn_terrain/}} to allow others to create their own instruments, rather than offering a static musical instrument with a fixed mode of interaction. In this way, the Latent Terrain system became a computational material, working together with autoencoders, customized audio corpora, and interfaces, in the ecology of musical interactions design. 


The intention of the project, therefore, became transforming the technical system into an accessible and easy-to-integrate material in DMI design. We packaged the mapping model into a collection of Max/MSP externals, offering a Max-style API that supports the workflow of model creation and inference. The API design was situated and guided by the first author's practice of DMI design and composing with autoencoders, which we reported in another publication \citep{zheng_exploring_2025}. In summary, we reflected on a long-term self-use of the system in a series of interface designs to identify the practical necessities in the workflow of integrating the latent space into DMI. The aim was to expose proper configurations to the API to allow for customizations and extended uses of the system, while maintaining simplicity to avoid technical knowledge becoming a barrier.

\begin{figure}[t]
\centering
\includegraphics[width=\linewidth]{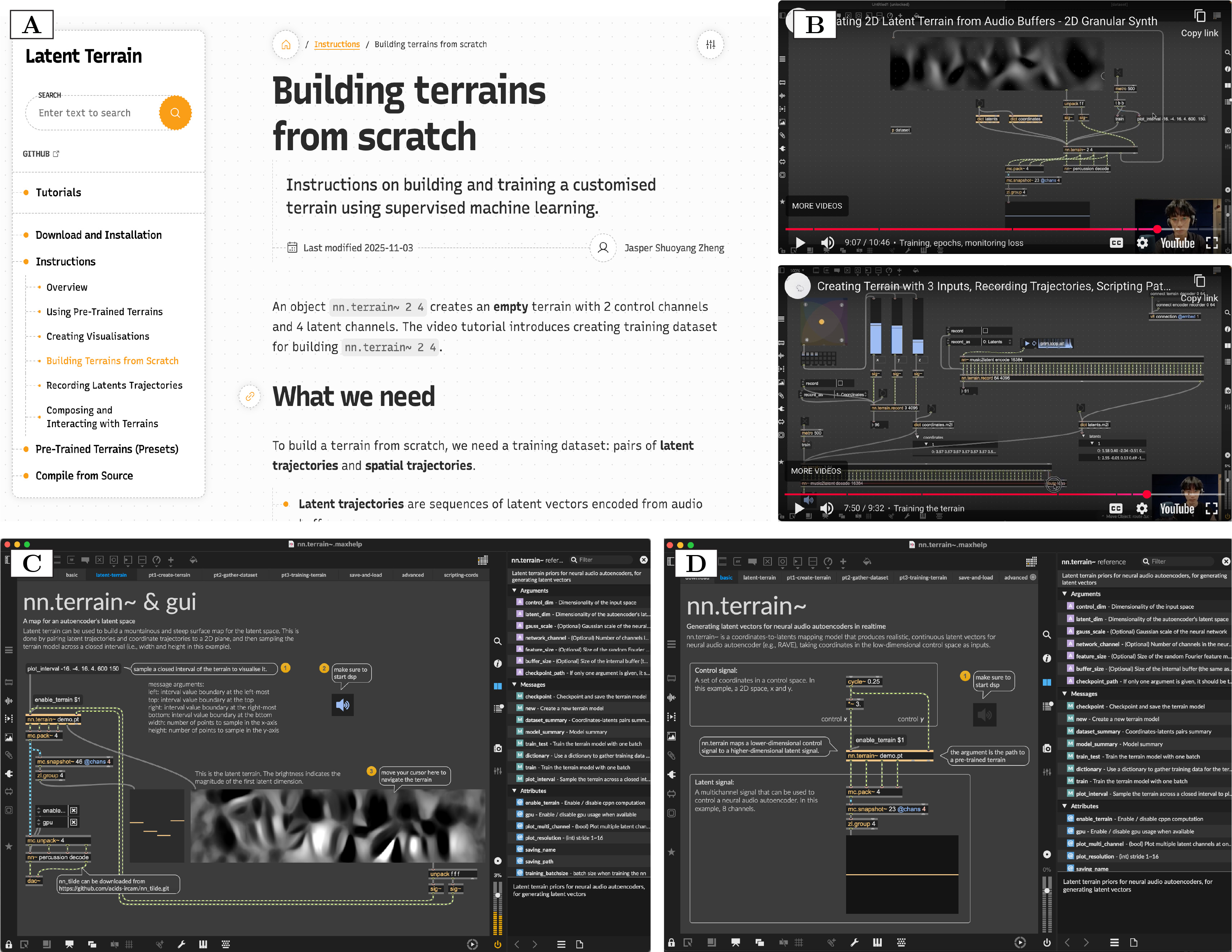}
\caption{A collection of materials to disseminate the research aspect of \textit{Latent Terrain} to hands-on experience in Max/MSP: (A) Online documentation, (B) video tutorials, (C and D) example patches as part of a getting-started guide.}
\label{fig_tutorial}
\end{figure}

To share our technical know-how gained throughout the development process, we maintain an online repository\footnote{\url{https://jasper-zheng.github.io/nn_terrain/}} of documentation and tutorials as a form of alternative research outcomes \citep{yoo_translating_2025}, shown in Figure~\ref{fig_tutorial}. The intended audience is audio programmers, musicians, DMI makers, and broader communities of creative practitioners. The online documentation provides an introduction to the package and a getting-started guide, using inclusive language to explain the inner architecture of autoencoders and entry points to unpack them. It also contains step-by-step instructions with video tutorials to cover a range of use cases of the package, from basic workflows such as loading and visualizing pre-trained mapping models, to in-depth customization such as creating one's own mapping model and experimenting with other input devices. 
On the Max/MSP side, each object in the package offers a help file containing a walkthrough of its functionalities, explanations of the API and configurations, and working examples of how it can be used in a patch.

We also maintain an open repository of autoencoders to enrich the pool of materials for the designers (links to these resources can be found on our online documentation page). This is motivated by the gap between research on autoencoders and their practical availability in musical platforms. The Realtime Audio Variational autoEncoder (RAVE) \citep{caillon_rave_2021} model is widely adopted by practitioners in NIME due to its availability to be used with the \texttt{nn$\sim$}\footnote{\url{https://github.com/acids-ircam/nn_tilde}} external in Max/MSP, and several other creative coding platforms, for real-time inference. However, a range of other neural audio autoencoders is exclusive to research environments, and can lack the consideration of real-time streamability at the development stage \citep{caillon_streamable_2022}. We added two other open-source autoencoders (Music2Latent \citep{pasini_music2latent_2024} and the autoencoder in the Stable Audio Open 1.0 model \citep{evans_stable_2025}) that vary in configurations and model architecture, and implemented them to the \texttt{nn$\sim$} external. We also made available a collection of RAVE models trained by the first author using Creative Commons datasets. This repository of resources aims to encourage a plural repertoire of materials, and to bridge the capability of recent research in neural audio synthesis to creative coding environments.


\subsection{Community Building with Artists}
\label{sec_projects}

To further encourage artistic exploration of the repository we created, we took a community-driven approach of allowing artists, designers, and digital luthiers to share their projects and technical know-how. The benefit of communities that intersect researchers, academic knowledge, and audiences has been recognised in various art and technology projects, such as the interactive machine learning tool Wekinator \citep{fiebrink_reflections_2020}, the corpora manipulation toolkit FluCoMa \citep{tremblay_enabling_2021, Tremblay_2022}. Inspired by these works, we create a space on our website \footnote{\url{https://jasper-zheng.github.io/nn_terrain/posts/}} for blog posts and project showcases, and invited a group of artists to explore a key question of ``how do neural audio models open a domain of design in musical interfaces''. 

\begin{figure}[t]
  \centering
  \includegraphics[width=\linewidth]{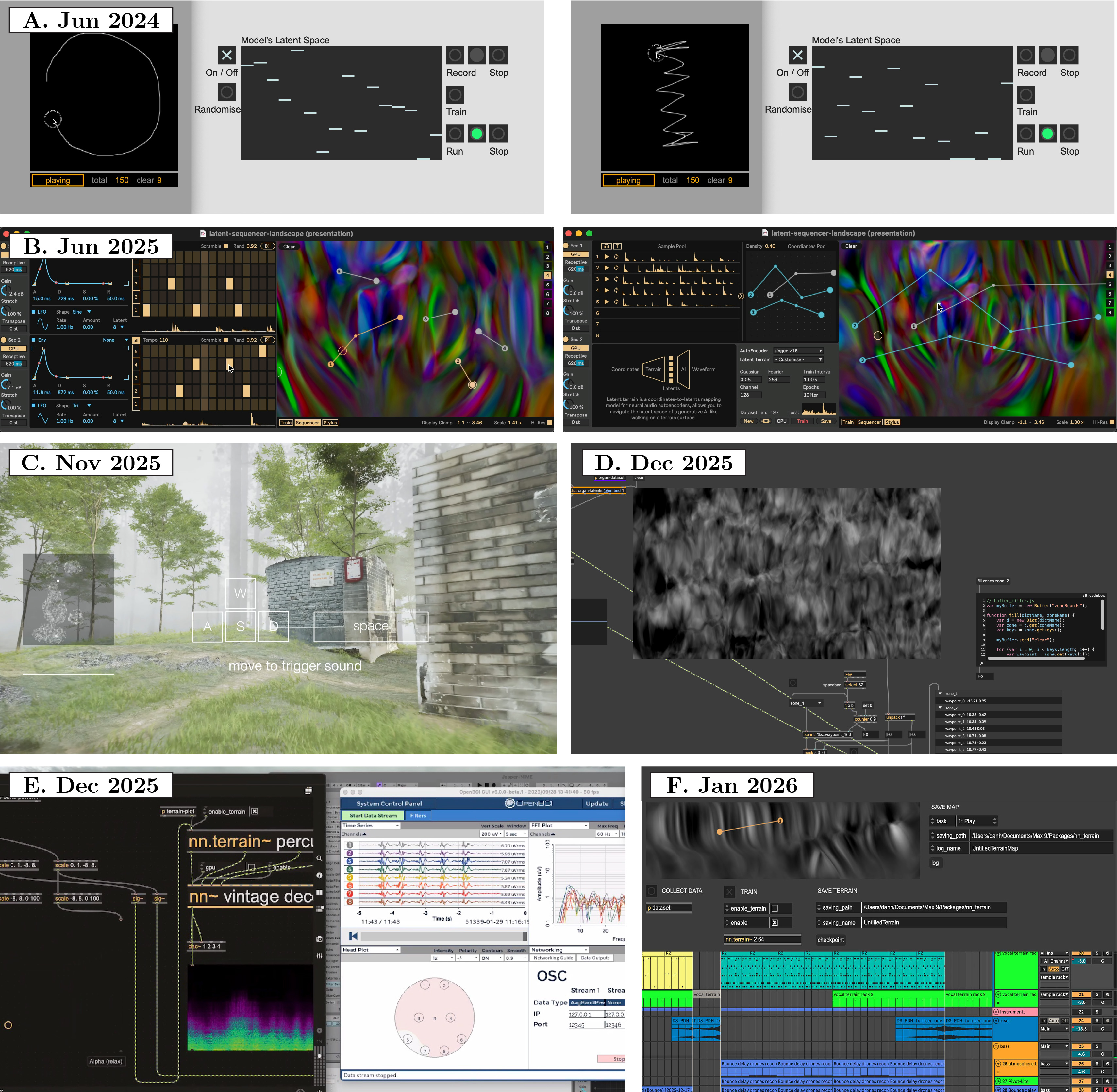}
  \caption{A collection of projects made using the Latent Terrain toolkit: (A) A MaxMSP interface for using visual sketches to steer a RAVE model; (B) A step sequencer that plays short ``latent samples'' while allowing users to adapt the terrain sound space on the fly. (C) \textit{nn/mémoire by Jiatong Liu}, an audio-visual virtual gallery with a theme of capturing the sound of cultural heritage in Hutongs (traditional northern Chinese courtyard houses in Beijing); (D) \textit{Trek by Nico García-Peguinho and Nikhil Bullock}, a collaborative composition and live performance system exploring sampling and interpolation strategies for the latent space; (E) \textit{Repressive Terrain by Keigo Yoshida}, an experimental data sonification patch with a theme of active interventions in a focused meditation; (F) \textit{ambient\_terrain\_1 by Dan Hearn}, an ambient soundtrack exploring the combination of AI and non-AI materials in the sample-based music production workflow.}
  \label{fig_projects}
\end{figure}

Figure~\ref{fig_projects} shows the collection of works in the repository. The first two projects (A and B), created by the authors as an initiation to explore the design space. The other four projects (C to F) are contributed by four artists who shared a mutual interest in exploring neural audio synthesis. Each project in the collection took a different approach to musical interface design, ranging from an immersive virtual gallery, a live performance, a brain–computer interaction, and an ambient soundscape composition.

Here we share reflections from the four artists to gain insight into their approach to understanding the models during their creation process. Quotes are taken from our conversations with each artist during the creation of the collection.

Various artists acknowledged the uniqueness of working with neural audio models by comparing them to conventional sound synthesis algorithms. For instance, Jiatong Liu, a creative technologist in machine learning, composition, and film, the creator of \textit{nn/mémoire}, made a comparison of neural audio models and conventional music production tools such as reverb or EQ: ``[neural audio models] shape a sound by entirely changing the structure and order of it''. Keigo Yoshida, an artist and scientist, the creator of \textit{Repressive Terrain}, emphasizes the interactive adaptation of sound spaces as an important iterative way of customising and working with the AI material. He made an analogy of adding training data to the Latent Terrain system with adding materials to an audio sampler, but ``it is able to shift timbre and musical structure across genres''. Nico García-Peguinho, an artist and researcher in machine listening and algorithmic composition, the co-creator of \textit{Trek}, highlighted that ``[neural audio synthesis] feels less intuitive than conventional oscillator-based synthesis''. 

Artists also mention the importance of learning from trials and errors. Dan Hearn, a creative technologist and musician, the creator of \textit{ambient\_terrain\_1}, describes his workflow with autoencoders for soundscape composition as ``explorative and serendipitous,’’ emphasising that the autoencoder ``gradually finds its place’’ within his existing sample-based practice through repeated trials, rather than being adopted as a standalone instrument with fixed affordances. Jiatong describes how working with neural audio models demands a shift in composition approach: ``need to learn to deal with the unpredictability through trial and error, without fully eliminating uncertainty''. Nico also highlighted that a tacit understanding of how to work with the material was made legible through active and embodied listening in the sound space. Keigo described the ``tracing the map like a musical score’’ as a compositional method that is difficult to articulate without hands-on engagement. 

There is also a demand for tools and facilities to support the hands-on engagement with neural audio models. For instance, Nico mentioned that conventional sound synthesis methods, such as sample-based synthesis or oscillator-based synthesis, benefit from a rich set of tools such as visualisation or preset saving and loading, whereas for neural audio models, ``very often you have to build things yourself, which makes them harder to work with''.

\section{Enabling Artists' Exploration of AI Materials}

Based on the development process of the Latent Terrain package, the online resources, and the reflections from the artists, here we discuss three takeaways on how to transform AI models into accessible and inclusive XAI materials in the crafting space of NIME, to inspire future XAI design and development.

\textbf{First, make visible how AI models can be tailored into the material assemblage of different artists' practices.} In NIME design, designers work with a range of synthesis algorithms, sensors, and the listening and composing practice, forming an \textit{assemblage} \citep{gioti_composing_2022}. Composing and weaving together a set of co-functioning materials in the assemblage is through the rapid cycles of trial and adjustment. Reflecting on the four creative projects, this iterative way of exploring and integrating materials is enabled by conditions in which they can probe, tune, and configure the AI model.  In this way, the role of autoencoders in creative practices became clear through the connection and interaction with other materials in the design space, rather than through predetermined affordances. In contrast to AI models that are constrained in technical research environments or API calls, the Latent Terrain toolkit unpacks the latent space into a creative coding environment, in which the autoencoder can belong to a broad constellation of AI and non-AI materials. The use and appropriation of AI as material can therefore be explored and understood by creative makers.

\textbf{Second, enable artists to learn technical and practical know-how of AI models through crafting.} DMI design, as a creative practice, is interdisciplinary technical work that involves apprenticeships and learning curves \citep{zayas-garin_dmi_2021}. Technical practice research is termed to relocate knowledge production in first-person, real-time technical practices \citep{pelinski_ways_2025}. This echoes the view of material-centered design, in which material understanding is built through close and iterative engagement, and enables skills of weaving together materials to fulfill a piece of design \citep{wiberg_material-centered_2018}. 

Explainability, in this respect, is not a one-time account of the model's inner work that will be disposed of immediately after the generation, but resources that can ensure makers' sustained practices \citep{clarke_longevity_2025} with the AI material. Within the four artistic pieces, makers do not shy away from tuning and troubleshooting their AI models. Instead, the complex and technical nature of latent space is acknowledged, but as described by Jiatong, they ``learn to deal with the unpredictability'', and by Nico, they ``seed help from the documentation to learn things like visualizing the latent parameters''. Crucially, this form of explainability is described as tacit understanding \citep{abuzuraiq_explainability--action_2025} that is uniquely attached between the maker and the material \citep{mcpherson_designing_2016}. This makes explainability inseparable from hands-on experimentation and listening to enable technique learning.

\textbf{Third, foster exchanges of domain-specific techniques and know-how among artists through community-building.} In creative practices, techniques and knowledge emerge when communities can repeatedly return to the same material to negotiate and collaborate \citep{saitis_timbre_2024}. This ocio-material environments \citep{noel-hirst_sampling_2025} is enabled by allowing practitioners to establish domain-specific vocabularies to share experiences and techniques. For instance, the account of the turntable is an example of how sustained practice and musical culture transform a technical artifact into a playable medium \citep{jorda_instruments_2004}. In the domain of AI for creativity, various works argue that ``providing the tools is not enough'' \citep{moore_making_2025}. Instead, the long-term and techno-fluent engagements with AI as material are enabled by shareable resources for learning and pedagogy \citep{fiebrink_meta-instrument_2009, moore_making_2025}. Within the practices with the Latent Terrain toolkit, artists discover their ways of knowing and working that can influence other practitioners. For instance, Nico's use of ``active listening'' method and Dan's use of ``latent resampling'' to characterize his workflow. This community aspect of explainability focuses on how the tacit knowledge of crafting and working can be articulated and circulated through communities to ensure a sustained practice with AI materials.


\section{Conclusion}

In this chapter, we suggest that XAI for creative practice can be approached as material explainability: a range of activities and artifacts that aim to bring AI models into the workspace as accessible and inclusive materials for design. Using latent space in neural audio models as a case study, we described how the Latent Terrain toolkit translates a technically complex and opaque representational space in AI models into a repository of software tools, documentation, resources, and tutorials for artists. Based on the repository and the reflection from a community of artists using it, we show how the material explainability approach supports hands-on artistic explorations of neural audio models in NIME design. We concluded by raising three suggestions on future XAI design for artistic practices: enable crafting and hacking by making visible how AI models can be tailored into different artists’ practices; enable artists to learn technical and practical know-how of AI models
through crafting; foster exchanges of domain-specific techniques and know-how among artists through community-building.

\begin{acknowledgement}
Shuoyang Zheng is a research student supported by the EPSRC UKRI Centre for Doctoral Training in Artificial Intelligence and Music (grant number EP/S022694/1). We thank all four artists for their engagement with the project and their contributions of their project descriptions, graphic content, and blog posts on the project webpage. 
\end{acknowledgement}

\ethics{Ethics Approval}{The project was approved by the Queen Mary University of London ethics committee, reference number: QMERC20.565.DSEECS25.026.}

\bibliographystyle{spmpsci-nodoi}
\bibliography{references-corrected.bib}

\end{document}